\documentclass[pre,twocolumn]{revtex4}
\usepackage{graphics}
\usepackage{amsmath}
\usepackage{amssymb}

\def\eqspace{\hspace{0.7cm}}
\def\PRE{{\it Phys. Rev. E} }

\def\half{1/2}

\def\JPA{{\it J. Phys A: Math. Gen.} }
\def\PRL{{\it Phys. Rev. Lett.} }
\def\smbox#1{\mbox{\tiny #1}}

\begin{document}
\title{Modelling one-dimensional driven diffusive systems by the Zero-Range Process}
\author{M. R. Evans$^1$, E. Levine$^2$, P. K. Mohanty$^2$, and D. Mukamel$^2$} \affiliation{$^1$ School of Physics, University of
Edinburgh, Mayfield Road, Edinburgh EH9 3JZ, United Kingdom \\
$^2$ Department of Physics of Complex Systems, Weizmann Institute
of Science, Rehovot, Israel 76100}
\date{4 May 2004}
%

\abstract{ The recently introduced correspondence between
one-dimensional two-species driven models and the Zero-Range
Process is extended to study the case where the densities of the
two species need not be equal. The correspondence is formulated
through the length dependence of the current emitted from a
particle domain. A direct numerical method for evaluating this
current is introduced, and used to test the assumptions underlying
this approach. In addition, a model for isolated domain dynamics
is introduced, which provides a simple way to calculate the
current also for the non-equal density case. This approach is
demonstrated and applied to a particular two-species model, where
a phase separation transition line is calculated. }
} 
\maketitle

\section{Introduction}
\label{intro} Phase separation in one-dimensional driven systems
has attracted much attention of late
\cite{Zia,Mukamel00,Evans00,Schutz03}.  In contrast to equilibrium
one-dimensional systems, where phase separation cannot occur
unless the interactions are long ranged, several examples of phase
transitions in one-dimensional non-equilibrium steady states have
been given
\cite{Evans98,LBR00,Rittenberg99,RSS,Kafri02A,Kafri03,CDE}. These
models generally have local noisy dynamics and some conserved
quantity or quantities driven through the system.

Particular attention has been paid to a simple yet general class
of models with two species of particles which are conserved under
the dynamics \cite{Zia,Schutz03}. These models are defined on a
ring, where each site can take one of three states: vacant,
occupied by a positive particle, or occupied by a negative
particle. Two conservation laws are obeyed by the dynamics, which
can be cast into two conserved quantities -- the total density of
particles in the system, $\rho$, and the fraction $\eta$ of
positive particles out of the total number of particles.

\begin{figure}
\centerline{\resizebox{0.4\textwidth}{!}{\includegraphics{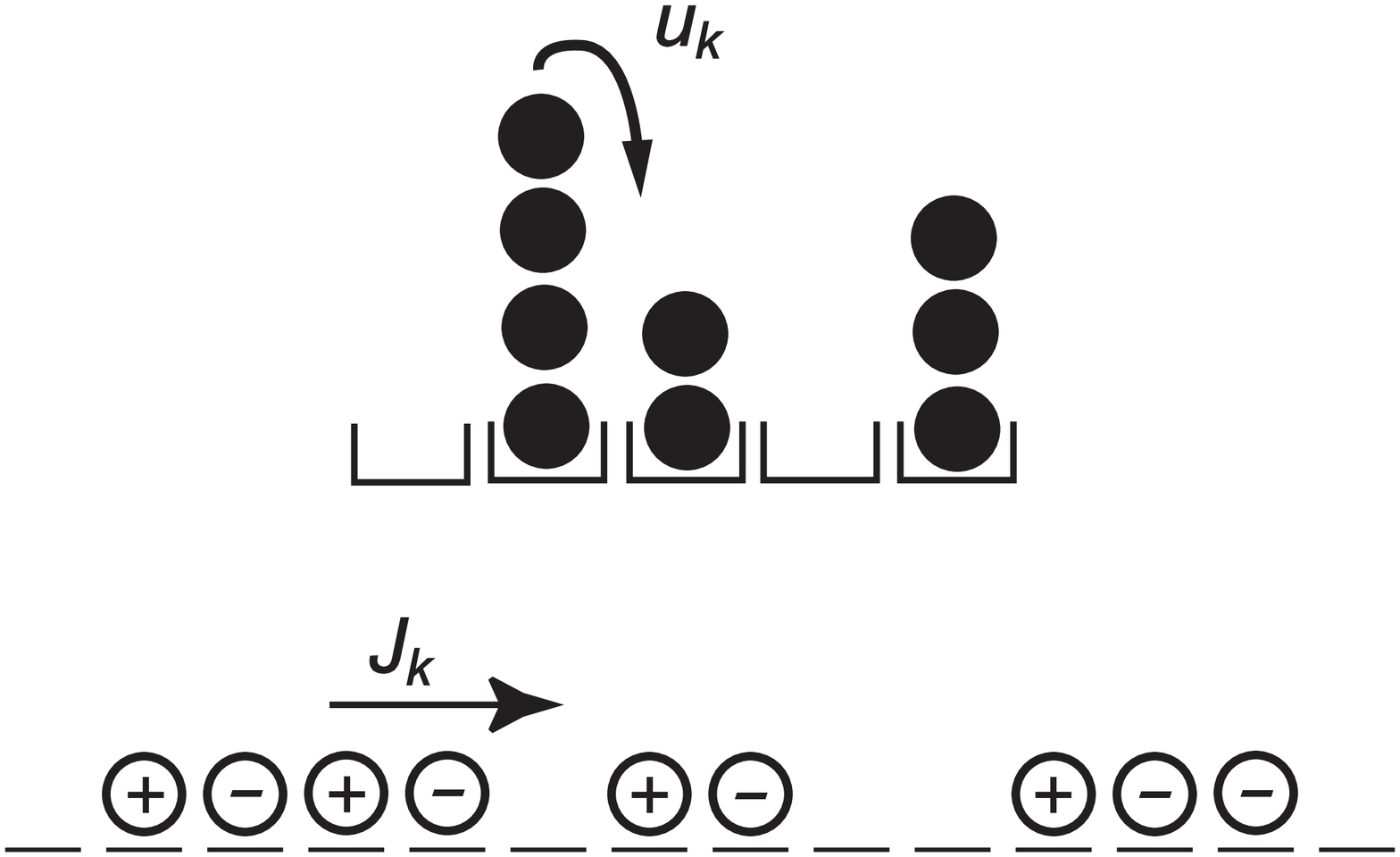}}}
\caption{A microscopic configuration of a two-species driven model
(bottom) and its corresponding configuration in the ZRP (top).}
\label{fig:cartoon}
\end{figure}

To study these models a coarse-grained description has been
developed \cite{Kafri02A}. In this description one views the
microscopic configuration of the model as a sequence of particle
domains, bounded by vacancies. Each domain is defined as a stretch
of particles of both types. The idea is to view particle domains
as urns which may exchange particles.  At a coarse-grained level
one identifies the current of particles through a domain as the
hopping rate of particles between neighbouring urns (see
Fig.~\ref{fig:cartoon}). This coarse-grained description then
defines a Zero-Range Process (ZRP) for which the steady state may
be solved exactly. When such correspondence is applicable one can
use the ZRP to obtain the distribution of the domain size,
although information about the correlation between the two species
of particles is lost.

Generally the identification of the driven system with a ZRP is at
a coarse-grained level and is not exact. Rather, it relies on the
applicability of some physical assumptions, as discussed below.
However, for a particular model for which an exact solution of the
steady state exists, it could be shown that the mapping of the
steady state to that of a ZRP is indeed exact \cite{Kafri02A}.

The correspondence between driven models and the ZRP may be used
to address the question of existence of phase separation in the
driven model \cite{Kafri02A}. In such a transition a fluid phase,
where the distribution of particles and vacancies is homogeneous,
becomes upon increasing $\rho$ a phase separated state. This state
is characterized by macroscopic particle domain devoid of
vacancies. In the context of the ZRP the transition into the phase
separated state corresponds to a condensation transition whereby
one urn becomes filled by a macroscopic number of particles
\cite{OEC,Evans00}.

A careful analysis of the corresponding ZRP suggested the
following criterion for phase separation. Let $u_k$ be the rate at
which particles leave a domain of size $k$. If $u_k$ vanishes in
the thermodynamic limit $k \to \infty$, phase separation is
expected at any density. Otherwise, for large domains, $u_k$
typically takes the form
\begin{equation}
\label{eq:current} u_k = u_\infty\left(1 + \frac{b}{k}\right)\;,
\end{equation}
where $u_\infty$ and $b$ are constants. The existence of phase
separation is then related to the value of $b$. Phase separation
cannot exist as long as $b\leq2$. However, if $b>2$ a phase
transition into a phase separated state occurs as the particle
density is increased. This picture has been used to argue against
the existence of phase separation in some models \cite{Kafri02A},
to study crossover phenomena \cite{Kafri02B}, and to suggest a
model which does exhibit a phase transition into a phase separated
state \cite{Kafri03}.

In order to apply the ZRP picture for a given model one has to
evaluate the rates $u_k$ by which domains exchange particles. This
may be  too difficult  a task to carry out analytically, and may
involve exceedingly large computation time to estimate
numerically. However, it has been suggested that in order to
estimate $u_k$ one may reduce the full many-domain system into a
single isolated domain problem. This is done by modelling an
isolated domain by an open chain of particles exchanging particles
with reservoirs at its ends. At the boundaries, particles are
injected and ejected at constant rates. When these rates are large
enough, the system is known to be forced into a maximal-current
phase with a bulk density $\eta=1/2$ \cite{Krug91}. This approach
is therefore only applicable when the densities of the two species
of particles in the full model are equal, and each domain is
stationary on average. In this case
the rate $u_k$ is directly related to the current $J_k$ which
flows through the domain. Thus, for equal densities of positive
and negative particles the many-domain problem is simplified to
the problem of a single domain of fixed size. The calculation of
the current through the single domain may be tackled numerically
or analytically, when possible
\cite{Kafri02A,Kafri03}.

It is important to notice that the approach outlined above for
describing a many domain system by a ZRP relies on two
assumptions. Firstly, it is assumed that domains are uncorrelated,
in the sense that the current $J_k$ flowing through a given domain
depends only on its own size.  Secondly, the current through a
domain of length $k$ is assumed to take its steady state value
with respect to a system of size $k$ (even though $k$ fluctuates)
and this steady state value is identified with that of an open
system.

In this paper we seek to test these underlying assumptions.
Whereas in previous studies the current $J_k$ of a domain of
length $k$ was studied using an isolated single domain, here we
introduce a numerical method for directly measuring $J_k$ of a
fluctuating domain within the full system. An agreement between
the two methods validates the assumptions behind the criterion.

We also seek to extend the approach to treat the case of non-equal
densities.  As discussed above, modelling a single domain by
an open system is not applicable in this case.
Moreover since we have unequal densities of positive and negative particles,
the vacancies  are not stationary and drift on average.
Thus in order to
reduce the many-domain problem to a single domain problem we have
to deduce the appropriate ensemble for the single domain.  In this
paper we propose a model for a fluctuating isolated single domain
which can be used to calculate a steady state current $J_k$ in a
domain of non-equal densities. This result is checked by using the
direct measurement discussed in the previous paragraph.
Moreoever in a limiting case we can solve the model exactly
through a matrix product ansatz and show that it produces the correct ensemble.

This paper is organized as follows. In Section~2 we define a
specific model that we use to demonstrate our approach. In
Section~3 we describe the method for direct measurement of the
currents within the many-domain system, and apply it to both cases
of equal and non-equal densities. A model for an isolated domain
which is not restricted to the case of equal densities is
introduced in Section~4 and an exact solution of a limiting case is given. We then discuss generalizations of the
approach to other models in Section~5, and present a summary and
outlook in Section~6.

\section{Model definition and some known results}
\label{sec:model}


In order to study the correspondence between driven diffusive
systems and the ZRP, we consider in the main body of this paper a
particular driven model as a test case. In this section we define
the model and present some analytical results.

The model is defined on a one-dimensional ring of $L$ sites. Each
site $i$ is associated with a `spin' variable $s_i$. A site can
either be vacant ($s_i=0$) or occupied by a positive ($s_i=+1$) or
a negative ($s_i=-1$) particle. Particles are subject to hard-core
repulsion and a nearest-neighbor `ferromagnetic' interaction,
defined by the potential
\begin{equation}
V = - \frac{\epsilon}{4} \sum_i s_i s_{i+1}\;.
\end{equation}
Here $0 \leq \epsilon < 1$ is the interaction strength, and the
summation runs over all lattice sites. The model evolves according
to the nearest-neighbour exchange rates
\begin{equation}
\label{eq:rates} +- \mathop{\longrightarrow}^{1+{\rm\Delta} V} -+
\eqspace +0 \mathop{\longrightarrow}^\alpha 0+ \eqspace 0-
\mathop{\longrightarrow}^\alpha -0\;,
\end{equation}
where ${\rm\Delta} V$ is the difference in the potential $V$
between the initial and final states. The number of particles of
each species, $N_+$ and $N_-$, are conserved by the dynamics.
Alternatively, the system can be characterized by two conserved
densities, namely the total density $\rho = (N_++N_-)/L$, and the
relative density $\eta = N_+/(N_++N_-)$ . This model, which is a
generalization of the Katz-Lebowitz-Spohn model
\cite{KLS,Hager01}, was introduced in \cite{Kafri03} and studied
for the case of equal densities of positive and negative
particles, $\eta = 1/2$.

\subsection{The non-interacting case, $\epsilon=0$}
\label{sec:model.e0} Let us first discuss the case $\epsilon = 0$,
where particles only interact through the hard-core exclusion. In this
case an exact solution shows that within a grand-canonical ensemble to
be defined below domains are uncorrelated, and the steady-state weight
factorizes into a product of single-domain terms. This is the case for
both equal and non-equal densities of the two species. These results
are obtained by considering a grand-canonical ensemble in which the
number of vacancies $M$ is kept constant. The number of particles of
each species, and thus the size of the lattice, are allowed to
fluctuate. A fugacity $\xi$ is attached to the positive particles,
thus controlling the relative density between the two species ($\xi=1$
corresponds to equal densities). All configurations of the system can
be described in terms of domains of particles, where a domain is
defined as an uninterrupted sequence of particles of both species. The
weight $W_M(\lbrace k_i \rbrace)$ of all configurations in which $k_i$
particles reside in the $i$th domain is then given by
\begin{equation}
W_M( \lbrace k_i \rbrace)=\prod_{i=1}^{M}{z^{k_i}Z_{k_i}(\xi)}\;,
\label{eq:mweight}
\end{equation}
where $Z_k$ is the sum over all weights of microscopic
configurations of a domain of length $k$, and $z$ is the fugacity,
which controls the overall density of particles in the system.
Hence, within the grand canonical ensemble domains are
statistically independent with a domain size distribution
\begin{equation}
P(k) \sim z^k Z_k(\xi)\;. \label{eq:distfun}
\end{equation}
The grand canonical partition function is given by
\begin{equation}
{\cal Z}_M = \sum_{\lbrace k_i
\rbrace}\prod_{i=1}^{M}{z^{k_i}Z_{k_i}(\xi)}\;. \label{eq:ZGC}
\end{equation}

In the equal density case the exact solution reveals that $Z_k$ is
identical to the partition function of the totally asymmetric
exclusion process on a one-dimensional lattice of $k$ sites with
open boundary conditions \cite{Derrida93}. For non-equal
densities, it turns out that $Z_k$ is identical to the
grand-canonical partition function of the same process defined on
a ring of size $k+1$ with a single vacancy
\cite{Mallick96,Sasamoto00}. In both cases
\begin{equation}
\label{eq:pfun} Z_k \sim (1+\sqrt{\xi})^{2k}/k^{3/2}
\end{equation}
for large $k$. The resulting distribution function
\eqref{eq:distfun} implies that the model does not exhibit phase
separation at any density. In particular, for any choice of $\rho$
one can choose the fugacity $z<1$ such that the average density
satisfies $\rho/(1-\rho) = \int k P(k) dk$.

Moreover, in the case $\epsilon=0$ the correspondence between the
steady-state of the model and that of the ZRP can be made
explicit. Within a grand-canonical ensemble of the ZRP urns are
statistically independent. The distribution function
$P_{\smbox{ZRP}}(k)$ for the occupation of a single urn is given
by $P_{\smbox{ZRP}}(k) \sim z^k \prod_{m=1}^{k}1/u_m$. On the
other hand, in the $\epsilon=0$ driven model the steady-state
current $J_k$ flowing through a domain is given by $J_k =
Z_{k-1}/Z_k$. The steady-state distribution function
(\ref{eq:distfun}) can then be
written as $P(k) \sim z^k \prod_{m=1}^k 1/J_m$. Thus
$P(k)=P_{\smbox{ZRP}}(k)$ with $u_k = J_k$. Using \eqref{eq:pfun}
one obtains $b=3/2$ for this case, implying no phase separation.

\subsection{The interacting case, $\epsilon > 0$, at $\eta=1/2$}
\label{sec:model.eps}  We now turn to the more general case,
$\epsilon \neq 0$. Here no exact mapping to the ZRP is available.
However, it was conjectured in \cite{Kafri02A,Kafri03} that the
physical picture obtained for the non-interacting case remains
valid, namely that the hopping rates of a corresponding ZRP should
be identified as the steady-state currents of isolated domains.
One therefore needs to calculate the asymptotic form of the
steady-state current running through a domain, $J_k(\epsilon) \sim
J_\infty(\epsilon)\left(1+b(\epsilon)/k\right)$. For $\eta=1/2$,
where the average domain velocity vanishes, the current $J_k$ may
be calculated by considering an isolated domain with open
boundaries, which exchanges particles with reservoirs at its ends
at high rates.

It has been argued \cite{Krug90,Krug97} that the coefficient
$b(\epsilon)$ of an isolated open domain is given by
\begin{equation}
b(\epsilon) = c\;b_R(\epsilon,\eta=1/2)\;. \label{eq:bs}
\end{equation}
Here $b_R(\epsilon,\eta)$ is the coefficient corresponding to a
closed fully-occupied ring, and $c$ is a universal constant which
is equal to $3/2$. For a ring the coefficient $b_R(\epsilon,\eta)$
can be calculated at any density $\eta$ \cite{Kafri03}. It is
given by
\begin{equation}
\label{eq:bofe} b_R(\epsilon,\eta) = - \frac{\lambda(\epsilon,
\eta) \kappa (\epsilon,\eta)}{2 J_\infty(\epsilon, \eta)}\;.
\end{equation}
In this expression \vspace{2pt} $\lambda=\partial^2 J^R_\infty /
\partial \eta^2$. The compressibility $\kappa =\lim_{k \to \infty} k^{-1}
\left(\langle \eta^2 \rangle - \langle \eta \rangle^2\right)$ is
evaluated in a grand-canonical ensemble of a fully-occupied ring
with average density $\langle\eta\rangle=1/2$. Using the known
properties of the steady state of this model \cite{KLS,Hager01} it
can be shown that
\begin{eqnarray}
\label{eq:Jofe} J^R_\infty(\epsilon,\eta) &=& \left[\gamma +
\epsilon \sqrt{4
\eta(1-\eta)}\right]\gamma^{-3} \\
\label{eq:kappa} \kappa(\epsilon,\eta) &=& \eta \left( 1-\eta
\right) \sqrt {1+4\eta \left( 1-\eta \right) \left( {\frac
{1+\epsilon}{1-\epsilon}}-1 \right) }
\end{eqnarray}
where $ \gamma =  [4 \eta \left( 1-\eta \right)]^{-1/2}+[(4 \eta(
1-\eta))^{-1} -1+(1+\epsilon)/(1- \epsilon)]^{1/2}$. Inserting
(\ref{eq:Jofe}) and (\ref{eq:kappa}) into (\ref{eq:bofe}) with
$\eta=\half$ one obtains the coefficient $b$ of an open domain as
a function of $\epsilon$ for the equal density case. It is found
that $b>2$ for $\epsilon
> 4/5$. The criterion mentioned above therefore implies that in this model
phase separation takes place at high density $\rho$ for any
$\epsilon>4/5$. Note that although the expression of
$b_R(\epsilon,\eta)$ is valid for arbitrary $\eta$ in the ring
geometry, the resulting $b(\epsilon)$ (Eq. \ref{eq:bs}) is
relevant to a domain with open boundaries only at $\eta=1/2$. A
single domain model which is applicable for $\eta \neq 1/2$ will
be discussed in Section~\ref{sec:neq}.

\begin{figure}
\centerline{\resizebox{0.4\textwidth}{!}{\includegraphics{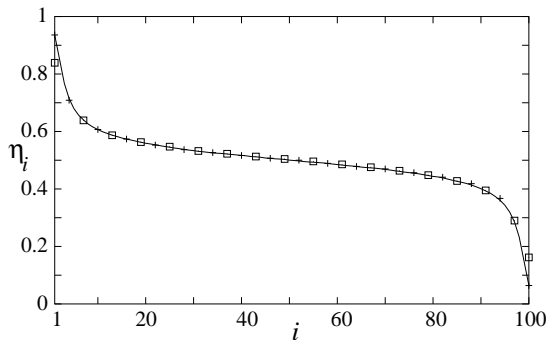}}}
\caption{Density profile of a domain of size $k=100$ at equal
densities ($\eta=1/2$) and $\epsilon=1/2$. Solid line corresponds
to direct simulation of the full many-domain model. Plus signs
correspond to an open isolated domain, and squares correspond to
the non-conserving single-vacancy ensemble.}
\label{fig:SymProfiles}
\end{figure}

\section{Direct numerical measurement of domain dynamics}
\label{sec:direct}

We now describe our method for a direct numerical measurement of
the flow of particles $u_k$ out of a domain of size $k$. As
discussed in the introduction, this method involves the full many
domain system, and thus allows a check on the validity of reducing
to the problem to that of a single domain.

The idea is during a simulation of duration $T$ to record the
number of hopping events out of a domain of size $k$, and the
average number of domains of size $k$. The ratio of these
quantities yields $u_k$. Here one unit of time corresponds to a
single Monte-Carlo sweep. We thus define
\begin{equation}
N_k = \frac{1}{T}\sum_{t=t_0}^{t_0+T} m_k(t) \eqspace F_k =
\frac{1}{T}\sum_{t=t_0}^{t_0+T} f_k(t)\;.
\end{equation}
Here $m_k$ is the number of domains of size $k$ residing in the
system at time $t$, and $f_k(t)$ is the number of exchanges $+0
\to 0+$ occurred between times $t$ and $t+1$ at the boundaries of
domains of size $k$. Of course, one can also define $f_k$ through
the transition rates $0- \to -0$, without changing the following
discussion. The measurement starts at time $t_0$, after short-time
relaxations are over. Clearly,
\begin{equation}
u_k = \lim_{T\to\infty} \frac{F_k}{N_k}\;.
\end{equation}
In practice, the  measurement time $T$ taken to be large enough to
ensure convergence. Estimates for $u_\infty$ and $b$ are then
obtained from the linear fit of $u_k$ to $1/k$. However one can
exploit the data obtained from numerical simulations better by
integrating the distributions. Thus we define
$$
\label{eq:tildes} \widetilde{N_k} = \sum_{\ell=k}^{\infty} N_\ell
~~~ \widetilde{F_k} = \sum_{\ell=k}^{\infty} F_\ell ~~~
\widetilde{Q_k} = \widetilde{N_k}^{-1}\sum_{\ell=k}^{\infty}
\frac{N_\ell}{\ell}\;.$$ Using $u_k = u_\infty(1+b/k)$ one has
\begin{equation}
\tilde{u}_k
=\frac{\widetilde{F_k}}{\widetilde{N_k}}=u_\infty\left(1+b
\widetilde{Q_k}\right)
\end{equation}
and one obtains $b$ and $u_\infty$ from a linear fit.

\subsection{Equal densities, $\eta = 1/2$}

While simulating the dynamics of the full model we have recorded
the density profiles of domains of a give size $k$. These profiles
are compared with those obtained from the single open domain
calculation in Fig.~\ref{fig:SymProfiles}. Both profiles are
identical to within the statistical fluctuations except at sites 1
and $k$ where there are systematic deviations. The deviations at
these two sites are to be expected as this is where the dynamics
for the single domain is simplified
 from the full model. The excellent agreement between the
profiles indicates that the single open domain properly models a
fluctuating domain in the full system.

\begin{figure}
\centerline{\resizebox{0.4\textwidth}{!}{\includegraphics{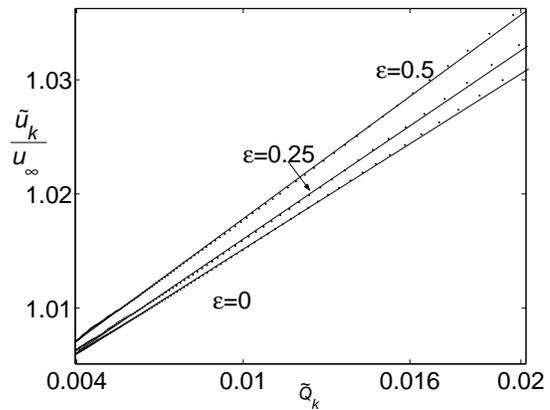}}}
\caption{Results of numerical simulations of the full model for
$\eta=1/2$ and $\rho=0.7$, from which $b(\epsilon)$ and $u_\infty$
are estimated. Solid lines are obtained by linear fits of the
data. }\label{fig:UofQ}
\end{figure}

We now apply the direct numerical method outlined above to the
case $\eta=1/2$. The results of numerical simulations are given in
Fig.~\ref{fig:UofQ} for several values of $\epsilon$.
Table~\ref{tab:e0} summarizes the resulting parameters
$b(\epsilon)$ and $u_\infty$. For $\epsilon=0$ one knows from the
exact correspondence of the model to ZRP that $u_k = J_k
=\frac{1}{4} \left(1+\frac{3/2}{k}\right)$. Our direct measurement
of $\tilde u_k$ recovers these  results quite faithfully. In
general, for $\epsilon>0$ we find that the measured values of $b$
and $u_\infty$ are in close agreement with
(\ref{eq:bs}--\ref{eq:kappa}). Thus we conclude that the current
flowing through a domain in the model can indeed be viewed as the
stationary current flowing trough an isolated open system of the
same size. The proposition that by increasing $\epsilon$ a phase
transition into a phase separated state occurs is thus verified.

\begin{table}
\begin{center}
\begin{tabular}{|c|c|c|c|c|}
\hline &$b(\epsilon)$& $b$& $u_\infty$ & $u_\infty$ \cr $\epsilon$
& (Eq. \ref{eq:bs})& measured &(Eq. \ref{eq:Jofe})  & measured\cr
\hline 0 &1.5 & 1.51& 0.25&0.25\cr 0.25 & 1.67 &1.65 &
0.2113&0.2115\cr 0.5& 1.82 & 1.79& 0.1585&0.1583\cr 0.8 &2 &
2.05&0.075&0.0749\cr \hline
\end{tabular}
\end{center}  \caption{Results of direct measurement of $u_k$ in the full model,
compared with analytical results of Section~2. We
estimate the error in the measurement of $u_\infty$ to be $0.0005$
and of $b$ to be $0.05$. \label{tab:e0}}
\end{table}

\subsection{Non-equal densities, $\eta \neq 1/2$}

We now consider the model with non-equal densities of the two
particle species, $\eta \neq 1/2$. In order to apply the criterion
for phase separation one needs to calculate the coefficient
$b(\epsilon,\eta)$. As mentioned in Section~\ref{sec:model}, an
exact mapping to the ZRP is only available at $\epsilon=0$, where
one finds $b=3/2$ for any value of $\eta$. To test the validity of
the correspondence to ZRP and evaluate $b(\epsilon,\eta)$ for
$\epsilon \neq 0$ and arbitrary $\eta$ one has to resort to direct
numerical simulations, as discussed above.

We have simulated the model for $\epsilon = 0.25, 0.5$ and
$\eta=0.6, 0.75, 0.85$. The values of $b$ are extracted from the
outflow of particles of domains of size up to $250$. We note that
we could obtain an accurate  estimate for $b$ in this way only for
the outflow of the majority species. Getting similar estimates for
the minority species would require significant statistics for much
larger domains.

\begin{figure}
\begin{center}\resizebox{0.5\textwidth}{!}{\includegraphics{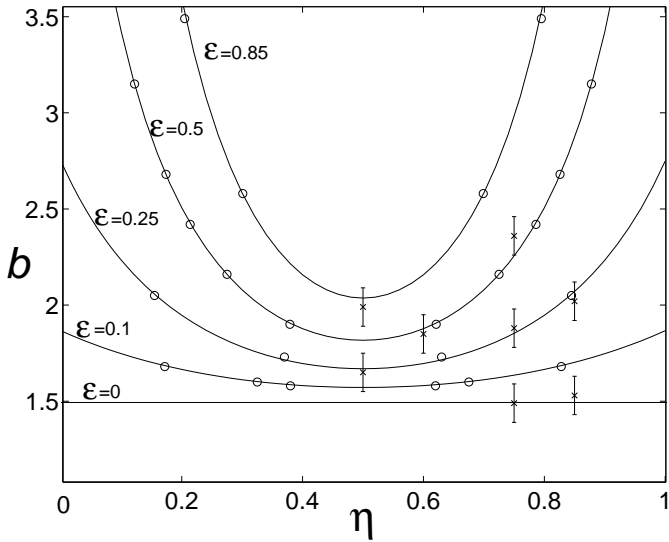}}
\end{center}\caption{The coefficient $b$ as a function of the
density $\eta$ of positive particles within a domain. Solid lines
are given by (\ref{eq:bs}), crosses are direct measurements in the
full system, while circles correspond to the non-conserving
vacancy ensemble. \label{fig:b}}
\end{figure}

In Fig.~\ref{fig:b} we display the values of $b$, as obtained from
direct measurement of the outflow of majority particles. Although
a priory one does not expect $b(\epsilon,\eta)$ to be given by the
expression obtained from the ring model of
Section~\ref{sec:model.eps}, $b(\epsilon,\eta) =
3/2\;b_R(\epsilon,\eta)$, we also display in this figure the
expression obtained from this formula. We find that the numerical
data agrees very well with these curves. This suggests that in
fact the analytical results obtained from the ring model
(\ref{eq:bs}--\ref{eq:kappa}) are valid for non-equal densities
($\eta \neq 1/2$) as well.


\begin{figure}
\begin{center}\resizebox{0.3\textwidth}{!}{\includegraphics{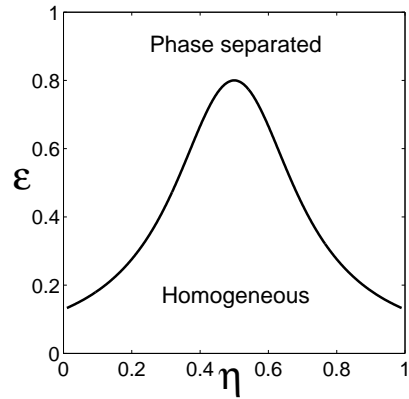}}\end{center}
\caption{Phase diagram for the model, as obtained by setting $b=2$
in Eq.~\ref{eq:bs}. \label{fig:pd}}
\end{figure}

We conclude that $b(\epsilon,\eta) = 3/2\;b_R(\epsilon,\eta)$,
where $b_R(\epsilon,\eta)$ is given by (\ref{eq:bofe}),
provides the correct expression for $b(\epsilon,\eta)$ for the
full model. According to the criterion discussed in the introduction,
a phase transition into a
phase separated state is expected at some critical density $\rho$
for $b>2$. The transition line in the $(\epsilon,\eta)$ plane
between the homogeneous and the phase separated state is depicted
in Fig.~\ref{fig:pd}. For values of $\epsilon$ which are larger
than $0.8$ phase separation is expected at high densities for any
value of $\eta$. On the other hand, for $\epsilon \lesssim 0.129$
phase separation does not occur at any density.

Direct numerical observation of the predicted transition line is
hard to obtain. For $\eta$ not too close to $1/2$ one would need
to simulate exceedingly large systems, far beyond our present
reach, in order for the fluid to sustain domains which are large
enough that the current flowing through them takes the
asymptotic form. It thus remains a challenge to devise some method
for numerical observation of this transition line.

\section{Single domain with non-equal densities: non-conserving vacancy ensemble}
\label{sec:neq} As discussed above, for the case of equal
densities of positive and negative particles one can model a
domain of length $k$ as an open boundary segment of length $k$
where particles enter and exit at the boundaries.  This is by
virtue of the fact that for sufficiently high entry and exit
rates, the open segment will be maintained in a maximal current
phase where the bulk density of positive particles organises
itself to be $\eta_{\rm bulk}= 1/2$. One uses the ensemble of the
open boundary problem to calculate the dependence of the current
on domain length.  The maximal current phase exhibits long-range
correlations that ultimately generate a slow decay of the current
with domain length and a condensation transition if $\epsilon$ is
sufficiently large.

However an open boundary segment cannot produce a bulk density
$\eta_{\rm bulk} \neq 1/2$ and retain long-range correlations i.e.
it cannot produce a maximal current phase with $\eta \neq 1/2$.
Thus, for the case of non-equal densities of positive and negative
particles one cannot use the ensemble generated by the open
boundary problem to calculate the current for a domain of size
$k$. Instead one must devise  an alternative ensemble that allows
the density of the domain to fluctuate about a value $\eta \neq
1/2$ whilst retaining the long-range correlations required in a
maximal current phase.

In this section we propose an ensemble for the single domain of
length $k$ that is generated by the dynamics of a single vacancy
on a ring of size $k+1$.
The same dynamics as the full model is used for the
exchange of particles (\ref{eq:rates}).
Also the dynamics of the vacancy retains
the processes present in the full model
\begin{eqnarray}
+\,0 &\rightarrow& 0\,+ \;, \qquad \textrm{with rate $\beta$}\;, \nonumber \\
0\,- &\rightarrow& -\,0 \;, \qquad \textrm{with rate $\alpha$}\;.
\label{alphbet}
\end{eqnarray}
However in addition we introduce two processes where particles are
not conserved
\begin{eqnarray}
+\,0 &\rightarrow& 0\,- \;, \qquad \textrm{with rate $\delta$}\;, \nonumber \\
0\,- &\rightarrow& +\,0 \;, \qquad \textrm{with rate $\gamma$}\;.
\end{eqnarray}
Note that in (\ref{alphbet}) we have generalised
(\ref{eq:rates}) to include a rate $\beta$.

We now show that in the case $\epsilon=0$ the above dynamics
generates precisely the ensemble required for a domain of length
$k$. To demonstrate this we solve exactly the steady state of the
non-conserving vacancy problem using a matrix product ansatz.

\subsection{Exact solution for $\epsilon = 0$}

The matrix product ansatz entails writing the steady state as a
product of matrices \cite{Derrida93,DJLS} i.e. the steady state
weight $w(\{ s_i \})$ for configuration $\{ s_i \} = s_1, \ldots,
s_{k}$. is
\begin{equation}
w(\{ s_i \}) = \rm{Tr} [X_1 \cdots X_{k}]\;,
\end{equation}
where the matrix ${\rm X}_i$ is
\begin{displaymath}
{\rm X}_i = \left\{
\begin{array}{ll}
D &\quad \textrm{if $\;s_i = +$}\;, \\
E & \quad\textrm{if $\;s_i = -$}\;, \\
A & \quad \textrm{if $\;s_i = 0$}\;.
\end{array} \right.
\end{displaymath}
Then it can be shown following \cite{Derrida93,DJLS}, that the
steady state weights for the present model can be written in this
form provided the matrices $D$, $E$ and $A$ satisfy the quadratic
relations
\begin{eqnarray}
DE  &=& D+\xi E \;,\label{DE}\\
\beta DA &=& \xi A \;, \label{DA}\\
\alpha AE &=& A \;, \label{AE}
\end{eqnarray}
where
$\xi \alpha \delta = \beta \gamma$.
For the conserving case $\gamma= \delta =0$, $\xi$ is not fixed
and may conveniently be set to $\xi =1$. This recovers the
previously known solution \cite{DJLS,Rittenberg99}. However, for
$\gamma$, $\delta \neq 0$, we must take
\begin{equation}
\xi = \frac{\beta \gamma}{\alpha \delta}\;. \label{eq:xi}
\end{equation}
Relations (\ref{DE}--\ref{AE}) are satisfied if we take $A$ to be
the projector $\vert V \rangle\langle W \vert$, where we employ a
bra-ket notation to denote the left and right vectors $\langle W
\vert$ and $\vert V \rangle$. Then letting $D = \xi \widetilde{D}$
relations (\ref{DE}--\ref{AE}) reduce to
\begin{eqnarray}
\widetilde{D}E  &=& \widetilde{D}+ E \;,\label{DE2}\\
\beta \widetilde{D}\vert V \rangle &=& \vert V \rangle \;, \label{DA2}\\
\alpha \langle W \vert E &=& \langle W \vert \;. \label{AE2}
\end{eqnarray}
Relations (\ref{DE2}--\ref{AE2}) obeyed by $\widetilde{D}$, $E$,
$\langle W \vert$ and $\vert V \rangle$, are precisely those
obeyed by the matrices and vectors used to solve the steady state
of the open boundary ASEP \cite{Derrida93}. Thus the weights for
the non-conserving vacancy are equal to those for an open boundary
system reweighted by a factor $\xi^{N_+}$; $\xi$ acts as a
fugacity to tune the relative density of positive and negative
particles.

The partition function for the non-conserving vacancy system is
given by summing over all possible configurations of positive and
negative particles on the ring and results in
\begin{equation}
Z_k = \mbox{Tr} \left[ A (\xi \widetilde{D} +E)^k \right] =
\langle W   \vert (\xi \widetilde{D} +E)^k    \vert V \rangle
\end{equation}
which is precisely the partition function for a domain of length
$k$ required in the grand canonical partition function for the
full system (i.e. fixed number of vacancies, fluctuating particle
numbers, see Section~\ref{sec:model}
(\ref{eq:mweight},\ref{eq:distfun},\ref{eq:ZGC}))
\begin{equation}
{\mathcal Z}_{M} = \left[\sum_{k=0}^\infty z^k
Z_k(\xi)\right]^M\;.
\end{equation}

Thus we have shown that in the case $\epsilon =0$ the
non-conserving vacancy generates the required ensemble for
domains. For $\epsilon >0$ we will provide numerical evidence that
this is still the case.

\subsection{Numerical Simulations for $\epsilon >0$}
\begin{figure}
\centerline{\resizebox{0.4\textwidth}{!}{\includegraphics{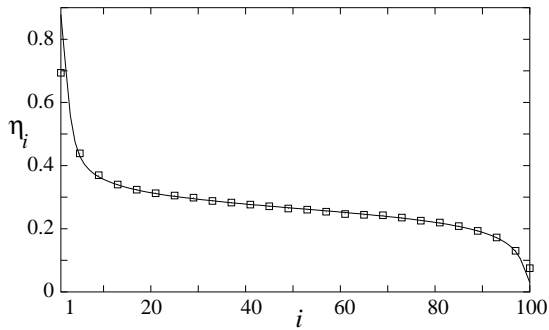}}}
\caption{Density profile of a domain of size $k=100$ at non-equal
densities $\eta=0.275$ and $\epsilon=1/2$. Solid line corresponds
to direct simulation of the full many-domain model. Squares
correspond to the non-conserving single-vacancy ensemble with $\xi
= 1/2$.} \label{fig:AsymProfiles}
\end{figure}

We ran numerical simulations of the non-conserving vacancy system
and measured  the density profiles as seen from the vacancy. In
Fig.~\ref{fig:SymProfiles} we compare the density profile for the
non-conserving vacancy problem at $\xi=1$ (corresponding to
$\langle\eta\rangle=1/2$) with the density profile of domains of
the same size computed in simulations of the full model and those
of an open isolated domain. Again we find the profiles identical
to within the statistical fluctuations except at sites 1 and $k$.
We have also compared the profiles obtained in the non-conserving
vacancy ensemble at non-equal densities ($\eta=0.275$) with those
obtained from the direct numerical simulations of the full model
(Fig.~\ref{fig:AsymProfiles}). The very close agreement of the
profiles in the bulk of the domains provide strong evidence that
the non-conserving vacancy generates the correct ensemble for
domains.

To evaluate $b(\epsilon,\eta)$ we measured the decay of the
current $J_k(\epsilon,\eta)$ with the system size $k$. As in
Section~\ref{sec:direct} we define the current as the rate at
which positive particles exchange with the vacancy. The current
$J_k(\epsilon,\eta)$ is obtained by simulating a system of size
$k+1$, fixing $\xi$ (Eq. \ref{eq:xi}) such that the average
density of positive particles matches $\eta$. The coefficient
$b(\epsilon,\eta)$ is then obtained by comparing the currents
$J_k$ with the asymptotic form $J_\infty\left(1+b/k\right)$. This
requires extensive numerical calculation since the current is
measured only through the single vacancy, taken to be at site 1,
and no spatial averaging takes place. To overcome this difficulty
simulations were performed using a multi-spin coding technique
\cite{multispin}. This allows many simulations to be run in
parallel utilising the same random numbers. The resulting
$b(\eta)$ for various values of $\epsilon$ is given in
Fig.~\ref{fig:b}. These results compare very well with the
analytical results obtained from the conserving ring model of
Section~\ref{sec:model.eps} (Eqs. \ref{eq:bs}--\ref{eq:kappa}),
and with the numerical results of the full model.

%
%

\section{Generalization to other models}

In previous sections we focused on the model presented in
Section~2. However, the direct numerical approach for calculating
$b$, introduced in Section~\ref{sec:direct} is applicable for
other models as well. This method may be used to test the
assumptions behind the correspondence of a driven model to the
ZRP, and the applicability of the criterion for phase separation,
discussed in the introduction.
Also, in cases where the correspondence between the driven model
and the ZRP is exact, the question of phase separation in the
driven model may be rigorously answered.

As an example where the numerical method for studying domains in
the full model we consider a two-lane variant of the model
discussed above. It has been proposed that when the rates
(\ref{eq:rates}) with $\epsilon=0$ are considered in a two-lane
geometry, phase separation may take place \cite{Korniss}. In this
case the model is defined on a lattice of size $2 \times L$, with
periodic boundary conditions in both directions, where particles
can move either within a lane or between the lanes. Direct
numerical simulations of the model suggest the existence of phase
separation at equal densities of the two species. However,
calculating $b$ by numerical simulations of single open domains
yields $b \simeq 0.8$, indicating that phase separation does not
occur in this model \cite{Kafri02A}. Since no exact mapping to ZRP
exists in this case, we have tested this result by carrying out
direct numerical measurement of $b$ in the full model as
introduced in Section~\ref{sec:direct}. We find that indeed $b
\simeq 0.8$ also in the full model, verifying the single open
domain approach in this case.

Finally, consider another variant of the model of Section~2,
whereby the particle exchange rate (\ref{eq:rates}) is replaced by
$+- \rightarrow -+$ with rate $1$ and $-+ \rightarrow +-$ with
rate $q$. This is a generalization of the $\epsilon=0$ case,
allowing for backward hopping. This model was introduced in
\cite{Rittenberg99} and studied in
\cite{Rittenberg99,RSS,Kafri02A,Kafri02B} for equal densities, and
in \cite{Arndt02} for non-equal densities. With $q<1$ this case is
qualitatively similar to the $\epsilon = 0$ case considered above,
with no phase separation taking place at any density. However, for
$q>1$ it can be shown that the steady state weight has the same
form as (\ref{eq:mweight}), with $Z_k \sim (1+\sqrt{\xi})^k
q^{\frac{1}{4}k^2}$ with $\xi=1$ for equal densities
\cite{Blythe00}. Since this model is exactly mapped onto a ZRP,
this result may be used to demonstrate that the model exhibits
phase separation at any non-vanishing densities $\rho$ and $\eta$.
From the above result for $Z_k$ it follows that for large $k$ the
current in this case is given by $J_k = Z_{k-1}/Z_{k} \sim
q^{-k/2}$, which vanishes in the limit $k \to \infty$. Thus the
criterion correctly predicts strong phase separation for all
densities of positive and negative particles.

\section{Summary and Discussion}
In this paper the correspondence between one-dimensional
two-species driven models and the Zero-Range Process is reviewed
and extended to consider the case of non-equal particle densities.
This is demonstrated for a two-species exclusion model with
`ferromagnetic' interactions. To apply this correspondence one has
to evaluate the length-dependence of the current emitted from a
domain of particles. Previous studies were restricted to equal
densities, were the average velocities of domains is zero. In
these studies domains were assumed to be statistically
independent, and the current is calculated using a model of single
domain with open boundaries. In the case of non-equal densities
domains have a non-zero average velocity, and thus this approach
is not applicable.

In the present work we have introduced a method for evaluating the
current of a domain of length $k$ by direct numerical simulation
of the full many-domain model. In the case of equal densities this
method yields the same results as before, verifying the validity
of the assumptions made in formulating the correspondence to ZRP.
Namely that domains are statistically independent and that the
current of a domain is given by the steady state current of an
isolated domain. Moreover, this method may be applied to the non-equal
density case.

We also introduced a model for a single domain, which enables one
to calculate the current of a domain without having to resort to a
simulation of the full model. Here a domain is modelled by a ring
with a single vacancy, with non-conserving dynamics at the vacant
site. It is demonstrated that this ensemble yields the same
density profiles and the currents as domains in the full model. It
thus provides a rather simple way of analyzing the full model.
Furthermore, we have outlined an exact solution for the non-conserving
vacancy model in the case $\epsilon =0$ which extends the range of models solved by the matrix product ansatz.

The phase diagram of the model in the interaction--density
$(\epsilon,\eta)$ plane has been calculated using both methods,
and the transition line to the phase separated state has been found.
Applications to other models have also been discussed.

We thank Yariv Kafri and Francesco Ginelli for helpful
discussions. This study was supported by the Israel Science
Foundation (ISF). Visits of MRE to Weizmann Institute were
supported by the Albert Einstein Minerva Center for Theoretical
Physics.

\end{document}